\documentclass[11pt, oneoside]{article}   	
\usepackage{geometry}                		
\usepackage{graphicx}				
\usepackage{authblk}
\RequirePackage{flushend}
\RequirePackage[numbers,sort&compress]{natbib}
\RequirePackage[colorlinks,citecolor=blue,urlcolor=blue,linkcolor=blue]{hyperref}
\RequirePackage{graphicx}
\RequirePackage{fix-cm}
\usepackage{dcolumn}
\usepackage{amssymb}
\usepackage{amsmath}
\usepackage{amsfonts} 
\usepackage{amsbsy}
\usepackage{color}
\usepackage{rotating}
\usepackage[english]{babel}
\usepackage{soul}
\usepackage{xcolor}
\usepackage{multirow}
\usepackage{hyperref}
\hypersetup{
colorlinks=true,
linkcolor=blue,
filecolor=magenta,
urlcolor=cyan,
citecolor=cyan
}

\newcommand{\be}{\begin{equation}}
\newcommand{\ee}{\end{equation}}
\newcommand{\bea}{\begin{eqnarray}}
\newcommand{\eea}{\end{eqnarray}}
\title{Equilibrium Temperature for Black Holes with Nonextensive Entropy}
\author[1]{Ilim \c{C}imdiker \footnote{ilim.cimdiker@phd.usz.edu.pl}}
\author[1, 2, 3]{Mariusz P. D\c{a}browski\footnote{mariusz.dabrowski@usz.edu.pl}}
\author[1]{Hussain Gohar\footnote{hussain.gohar@usz.edu.pl}}
\affil[1]{Institute of Physics, University of Szczecin, Wielkopolska 15, 70-451 Szczecin, Poland}
\affil[2]{National Centre for Nuclear Research, Andrzeja So{\l}tana 7, 05-400 Otwock, Poland}
\affil[3]{Copernicus Center for Interdisciplinary Studies, Szczepa\'nska 1/5, 31-011 Krak\'ow, Poland}
\date{}							
\begin{document}
\maketitle
\begin{abstract}
Hawking temperature has been widely utilised in the literature as the temperature that corresponds to various nonextensive entropies. In this study, we analyze the compatibility of the Hawking temperature with the nonextensive entropies. We demonstrate that, for every nonextensive entropy, one may define an effective temperature (which we call equilibrium temperature) by utilizing the equilibrium condition, and that there is always an additive equilibrium entropy associated with this effective temperature. Except for Bekenstein entropy, we show that Hawking temperature is thermodynamically inconsistent with other nonextensive entropies. We focus on the equilibrium requirement for the Tsallis-Cirto black hole entropy and demonstrate that the Bekenstein-Hawking entropy is the related equilibrium entropy, and the Hawking temperature is the associated equilibrium temperature for the Tsallis-Cirto black hole entropy. 

\end{abstract}

\section{Introduction}
The seminal works of Bekenstein \cite{Bekenstein:1973ur}, and Hawking \cite{Hawking:1974rv,Hawking:1976ra} on the thermodynamics of black holes \cite{Gibbons:1976pt,Hawkingnew1,Hawking:1982dh,Bardeen:1973gs} have a wide range of applications in gravitation and cosmology. These concepts have been applied, for instance, to investigate gravity from a thermodynamic perspective  \cite{Verlinde:2010hp}, to derive Einstein's field equations from the first law of thermodynamics \cite{Jacobson:1995ab}, to study holographic dark energy  \cite{Wang:2016och}, and to examine the universe's accelerated expansion from a thermodynamic perspective  \cite{Padmanabhan:2009kr, Cai:2010hk, Cai:2005ra, Easson:2010av}. Numerous studies have been made to extend these concepts from a thermodynamic and quantum perspective, including quantum gravity corrections \cite{Scardigli:1999jh,Alonso-Serrano:2018ycq, Meissner:2004ju, Rovelli:1996dv} and thermal fluctuation corrections \cite{Das:2001ic}. 
Quantum field theory is used to study Hawking radiation by incorporating quantum effects on the horizon \cite{Hawking:1974rv, Hawking:1976ra,Parikh:1999mf,Ejaz:2013fla}. This enables the calculation of the Hawking temperature, which supports Bekenstein's idea of a black hole's entropy. This concept of entropy is somewhat geometric and relies on Hawking's area theorem \cite{PhysRevLett.26.1344}. The laws of black hole thermodynamics \cite{Bardeen:1973gs} are analogous to the laws of classical thermodynamics by defining the Bekenstein-Hawking entropy $S_{bh}$ \cite{Bekenstein:1973ur} and the Hawking temperature $T_{bh}$ \cite{Hawking:1974rv} as \footnote{ We utilize natural units by taking the speed of light $c$, the Newton's constant $G$, the reduced Planck's constant $\hslash$, and the Boltzmann's constant $k_B$ equal to one. We have introduced $k_B$ in some sections to check the dimensional consistency.}
\be
S_{bh}=\frac{A}{4},~~~ T_{bh}=\frac{\kappa}{2\pi}. \label{S_bh}
\ee
For the case of a Schwarzschild black hole with mass $M$, the area $A$  becomes $A=4\pi r_h^2=16\pi M^2 $, where $r_h =2M$ is the Schwarzschild radius, and the surface gravity $\kappa$  becomes $\kappa=1/4M$.
By using the quantities in equation (\ref{S_bh}), the first law of black hole thermodynamics for Schwarzschild black hole can be written as
$dM =\kappa dA/8\pi,$ 
which is equivalent to the first law of thermodynamics $dE=TdS-PdV$, except for the pressure-volume term $PdV$ \cite{Dolan:2012jh}, with mass $M$ playing the role of internal energy $E$, $\kappa/2\pi$ playing the role of temperature and $A/4$ playing the role of entropy, respectively. The $PdV$ term can be introduced for an anti-de Sitter (AdS) black hole by considering the negative cosmological constant $\Lambda <0$ as pressure $P$ \cite{Kastor:2009wy,Kubiznak:2012wp,Kubiznak:2014zwa,Caldarelli:1999xj}, and introducing the volume $V$ as
\be
P=-\frac{\Lambda}{8\pi}, ~~~ V=\frac{4}{3}\pi r_h^3,
\ee \label{P}
so that the extended first law of thermodynamics reads as
\be
d M=T_{bh}d S_{bh}-P dV. \label{dM}
\ee
 In this way, the Smarr formula \cite{Smarr:1972kt} for the mass of the black hole reads 
\be
M=2T_{bh}S_{bh}-2PV. \label{M}
\ee
Taking the cosmological constant as a thermodynamic pressure provides the notion of volume for black holes, which is missing from the first law of black hole thermodynamics. In this scenario, the mass $M$ {\it no longer represents} the black hole's internal energy. However, it now acts similarly to the gravitational equivalent of enthalpy, which is the sum of internal energy $E$ and the work term $PV$. Furthermore, there are interesting consequences in this scenario \cite{Kubiznak:2014zwa,Wei:2014hba,Altamirano:2013ane,Altamirano:2013uqa}. For example, black holes act like Van der Waals fluids. In this manner, intriguing phase behavior such as the reentrant phase transition and triple points in the context of black holes have been examined.
For more details, see \cite{Kubiz_k_2017} and references therein. 

The primary difficulty with the idea of black hole entropy is the absence of a suitable statistical mechanical explanation. Instead, it must rely on Bekenstein's definition, which states that because black hole entropy is directly proportional to the area of its event horizon rather than its volume, it is nonadditive and presumed to be nonextensive. Because of this, black holes and other cosmological and gravitational applications cannot be well described by classical thermodynamics or statistical mechanics. Instead, many nonextensive statistical mechanics approaches have been used to examine black holes and other cosmological applications \cite{Nojiri:2021czz,Nojiri:2022aof,Nojiri:2022sfd,Nojiri:2021jxf,Promsiri:2020jga,Promsiri:2021hhv,Tannukij:2020njz,Nakarachinda:2021jxd,Promsiri:2022qin,Nakarachinda:2022gsb,Saridakis:2020zol,Dabrowski:2020atl,Nojiri:2022ljp,Komatsu:2016vof,Komatsu:2015nkb,Nunes:2015xsa,Liu:2022snq,Majhi:2017zao,Luciano:2021mto,DiGennaro:2022ykp,DiGennaro:2022grw,Asghari:2021bqa,Abreu:2022pil}. These nonextensive approaches include a variety of proposals. Tsallis statistics \cite{Tsallis:1987eu,tsallisbook}, R\'enyi statistics \cite{reny1}, the Tsallis-Cirto \cite{Tsallis:2012js,Tsallis:2019giw} and Barrow entropy \cite{Barrow:2020tzx} for the case of black holes, Sharma-Mittal statistics \cite{SM2,SM,SM1}, and Kaniadakis statistics \cite{KD} are a few examples. Long-range forces are taken into account in Tsallis' statistical mechanics or thermodynamics, and nonextensivity results from these long-range interactions, which are ignored in traditional Gibbs statistics. R\'enyi statistics are widely used in quantum information, and R\'enyi entropy measures the entanglement of quantum systems. Sharma-Mittal statistics is only an extension of Tsallis and R\'enyi statistics, whereas Kaniadakis statistics is inspired by the Lorentz transformation of special relativity. Barrow entropy is mathematically equal to Tsallis-Cirto entropy in the case of black holes, but the motivation for it derives from the fractal structure of the black hole horizon caused by quantum fluctuations. Tsallis-Cirto black hole entropy is driven to make the Bekenstein entropy additive and extensive. 

For the classical thermodynamic systems in thermal equilibrium, the zeroth law utilizes the transitivity relationship between systems to define an empirical temperature for each system. 
Another method of defining temperature is to use the equilibrium condition of the system, which maximizes its overall entropy. In addition to this, entropy and internal energy must adhere to the additive composition rule.
The additivity of entropy corresponds to ignoring long-range forces, whereas the additivity of internal energy refers to ignoring the interaction energy between composite systems. The temperatures derived from the transitivity relation and the equilibrium condition in traditional Gibbs thermodynamics are equal in this regard. 
Interestingly, considering a system's strong interaction with a thermal bath, such as for a strongly coupled system, the zeroth law's criterion is based on equilibrium conditions rather than transitivity relations \cite{Hsiang:2020hha}.
Therefore, there is a clear distinction between the terms ``in equilibrium" and ``in thermal equilibrium". It also implies that a system may approach equilibrium but not necessarily thermalize when it is strongly coupled to a bath. This means that the additive composition rule for both internal energy and entropy is sufficient for thermal equilibrium in Gibbs thermodynamics. 
 
In nonextensive thermodynamics, the zeroth law has numerous issues, as noted in \cite{PhysRevE.67.036114, OU2006525,Bir__2011}. Because long-range forces are considered, entropy does not obey the additive rule. We can assume that the internal energy is additive by considering weakly interacting systems. The equilibrium condition provides an effective equilibrium temperature \cite{Abe_2001}, from which one can calculate the equilibrium entropy for a system. This equilibrium temperature can be called a zeroth law compatible temperature in the nonextensive setup because the absolute temperatures defined for each subsystem differ from the equilibrium temperatures derived from the equilibrium condition. Like, as for the strongly coupled quantum systems, the nonextensive systems can be ``in equilibrium" at the effective temperature but not ``in thermal equilibrium". In fact, the zeroth law in a nonextensive setup can be only defined by using the equilibrium condition.  Similarly, the corresponding equilibrium entropy differs from the nonextensive one and follows the additive composition rule.

Nonextensive thermodynamics has several applications in cosmology and gravitation.
Most cosmological investigations use nonextensive entropies along with the Hawking temperature. The consistent thermodynamic quantities that follow the thermodynamic relations are essential to these applications from a thermodynamic point of view. For example, R\'enyi temperature and R\'enyi entropy are consistent thermodynamically when employed together, as in the case of Hawking temperature and Bekenstein entropy. The question now is whether nonextensive entropies defined on black holes and cosmological horizons are thermodynamically consistent with Hawking temperature. To answer this question, this article focuses on a few composition rules for different definitions of the nonextensive entropy of black holes. We will focus on the equilibrium conditions and examine the equilibrium temperatures in the nonextensive setup. We shall also comment on the thermodynamic consistency of Hawking temperature with nonextensive entropies by using standard thermodynamic relations. Additionally, we will utilize the Schwarzschild black hole as an example of a thermodynamic system. In this context, we will not explore any cosmological models, but the justification and analysis in this paper will hold for cosmological models as well.  

The structure of the paper is as follows. In section \ref{nonextensive}, we introduce the Tsallis nonextensive entropy and find the corresponding equilibrium temperature. In section \ref{RenyiETM}, we apply it to the Schwarzschild black hole and  define its energy, temperature, and mass. In section \ref{composition3}, we investigate the Tsallis-Cirto black hole entropy and equilibrium conditions and we briefly mention the relation between the Tsallis-Cirto and Barrow entropies. Finally, in section \ref{summary}, we summarize our main conclusions.

\section{Composition Rule and Equilibrium Temperature in Tsallis Nonextensive Setup}
\label{nonextensive}
Tsallis nonextensive entropy generalizes the Gibbs-Shannon's entropy into \cite{tsallisbook}
\be
S_q=-k_B\sum_{i}[p(i)]^q\ln_qp(i), \label{S_T}
\ee 
where $p(i)$ is the probability distribution defined on a set of microstates $\Omega$, the parameter $q$ determines the degree of nonextensivity and we consider it positive to ensure the concavity of $S_q$. The q-logarithmic function $\ln_qp$ is defined as
\be
\ln_qp=\frac{p^{1-q}-1}{1-q},
\ee 
such that, in the limit, $q \to 1$, the equation (\ref{S_T}) reduces to Gibbs-Shannon's entropy
\be
S_G=-k_B\sum_ip(i)\ln p(i). \label{SG}
\ee
Note that Tsallis entropy (\ref{S_T}) satisfies a nonadditive composition rule, which we shall discuss in the next section, while Gibbs entropy (\ref{SG}) satisfies the additive composition rule. However, via ``formal logarithm" approach \cite{Bir__2011}, one can write a corresponding additive entropy in terms of $S_q$ such that 
\be
S_R=\frac{k_B}{1-q}\left[\ln{\left(1+\frac{1-q}{k_B}S_q\right)}\right],
\label{ERenyi}
\ee
which happens to be the R\'enyi entropy \cite{reny1}
\be
S_R= k_B \frac{\ln \sum_i p^q(i)}{1-q}. \label{S_R}
\ee
Later, we shall see that $S_R$ is related to the equilibrium condition and it will be equilibrium entropy for a nonextensive system, which will also correspond to an equilibrium temperature defined from the equilibrium condition by maximizing the nonextensive entropy (\ref{S_T}).

By following \cite{Abe_2001c}, we consider a thermodynamic system composed of two independent subsystems, $1$ and $2$, in contact with each other. By defining a general composition rule
\be
S_{12}=f(S_1, S_2),
\ee 
which tells us that any total entropy $S_{12}$ can be expressed in terms of the entropies of subsystems $S_1$ and $S_2$. Here, $f$ is a bivariate function of the $C^2$, and it is assumed to be symmetric. In this context, the Gibbs-Shannon entropy $S_G$ satisfies 
\be
f(S_{G1}, S_{G2})=S_{G12}=S_{G1}+S_{G2},
\ee 
and 
Tsallis nonextensive entropy $S_q$ follows the following general nonadditive composition rule \cite{Abe_2001c}
\be
 S_{q12}=S_{q1}+S_{q2}+\frac{\lambda}{k_B}S_{q1}S_{q2}, \label{S_12}
\ee
for a thermodynamic system having total entropy $S_{q12}$, which is composed of two independent subsystems having entropies $S_{q1}$ and $S_{q2}$, which are in contact with each other thermally. Here, we introduced $\lambda=1-q$.  in above equation (\ref{S_12}).

Since we are interested in equilibrium thermodynamics, we consider fixed total internal energy $U_{q12}=U_{q1}+U_{q2}$ for a composite system, where $U_{q1}$ and $U_{q2}$ are the internal energies of the indivisible subsystems, and the equilibrium condition can be found by maximizing the equation (\ref{S_12}), i.e., $\delta S_{q12}=0$ with $\delta U_{q12}=0$, which gives the following condition
\be
\frac{k_B\frac{\partial S_{q1}}{\partial U_{q1}}}{1+(\lambda /k_B)S_{q1}}=\frac{k_B\frac{\partial S_{q2}}{\partial U_{q2}}}{1+(\lambda /k_B)S_{q2}}=k_B\beta^*, \label{beta*}
\ee
where $k_B\beta^*$ is a separation constant and we introduced $\beta$ for each subsystem, which is  defined as
\be
k_B \beta =\frac{\partial S_{q}}{\partial U_{q}}.
\ee 
\label{EC}
Now we can easily write down the effective temperature as an equilibrium temperature by using the equilibrium condition (\ref{beta*}) such that
\be
T_{eq}=\frac{1}{k_B\beta^*}=(1+\frac{\lambda}{k_B}S_q)\frac{1}{k_B\beta}. \label{Tphys} 
\ee
Similarly, the equilibrium pressure $P_{eq}$ can be defined in the state of mechanical equilibrium by maximizing the entropy (\ref{S_12}) with fixed total volume $V=V_1+V_2$ of the composite system and individual subsystem volumes $V_1$ and $V_2$, which gives the following condition
\be
\frac{\partial S_{q1}/\partial V_1}{1+(\lambda /k_B)S_{q1}}=\frac{\partial S_{q2}/\partial V_2}{1+(\lambda /k_B)S_{q2}}=\frac{P_{eq}}{T_{eq}},
\ee
so that the physical pressure reads as
\be
P_{eq}=\frac{T_{eq}}{1+(\lambda/k_B)S_q}\frac{\partial S_q}{\partial V}.
\ee
We shall see that the Clausius relation modifies due to the above equilibrium temperature and the equilibrium pressure.

In order to develop the nonextensive thermodynamic relations, we use the Legendre transformation and the first law of thermodynamics. In \cite{TSALLIS1998534}, the free energy $F_q$, as the Legendre transform structure, in the context of nonextensive thermodynamics, is defined as 
\be
F_q=U_q-\frac{1}{k_B \beta}S_q. \label{leg1}
\ee
In the above equation, the variable in front of $S_q$ is the inverse of Lagrange multiplier $\beta$ which gives the nonphysical temperature. However, all thermodynamic quantities should be written in physical variables. Therefore, in \cite{Abe_2001}, Abe et al. proposed the following generalized free energy
\be
F_q=U_q-T_{eq}\frac{k_B}{\lambda}\ln\left(1+\frac{\lambda}{k_B}S_q\right), \label{leg2}
 \ee
where $\beta^*(\beta)$ is introduced which gives the effective equilibrium temperature $T_{eq}$. In order to define the modified Clausius' relation, take the derivative of $F_q$ and using the first law of thermodynamics $dQ_q=dU_q+P_{eq}dV$, one can write
\be
\frac{k_B}{\lambda}d\ln(1+\frac{\lambda}{k_B}S_q)=\frac{dQ_q}{T_{eq}}.
\ee
which is modified by Clausius' relation for nonextensive systems. From the above equations (\ref{leg1}) and (\ref{leg2}), we can define a new form of entropy and we denote it by $S_{R}$
\be \label{S_R1}
S_{R}=\frac{k_B}{\lambda}\ln(1+\frac{\lambda}{k_B}S_q).
\ee
Furthermore, the new equilibrium entropy $S_{R}$, by using the composition rule (\ref{S_12}) of $S_q$, follows the additive rule, which can be easily shown as
\be
S_{R12}=S_{R1}+S_{R2}.
\ee
Interestingly, this new definition of entropy $S_R$ happens to be the R\'enyi entropy if the Tsallis entropy $S_q$ is given. 
\section{R\'enyi Black Hole Entropy, Temperature, and Mass}
\label{RenyiETM}
In this section, we discuss the application of the above equations  (\ref{Tphys}) and (\ref{S_R1}) to the Schwarzschild black hole.

We assume that Bekenstein-Hawking entropy (\ref{S_bh}) is the Tsallis entropy $S_q$ in (\ref{S_R1}) and write down the corresponding equilibrium entropy $S_R$ in terms of $S_{bh}$ and the equilibrium temperature $T_{eq}$ in terms of $T_{bh}$.

For the case of the Schwarzschild black hole\footnote{We do not consider other types of black holes here, but our formulism is general and can also be applied to them. For example, one can follow \cite{Hirunsirisawat:2022ovg}.} , we can write the R\'enyi entropy $S_{R}$ (\ref{ERenyi}) as
\be
S_{R}=\frac{1}{\lambda}\ln(1+\pi\lambda r_h^2), \label{SRbh}
\ee
where we have used $k_B=1$. The equilibrium temperature $T_{eq}$ (\ref{Tphys}) as (cf. formula (19) of Ref. \cite{Alonso-Serrano:2020hpb})
\be
T_{eq}=T_{R}=\frac{1}{4\pi r_h}+\frac{\lambda r_h}{4}.\label{TRbh}
\ee
By using the equations (\ref{SRbh}) and (\ref{TRbh}), we write the mass $M_R$ of the R\'enyi black hole by using the relation for the Smarr mass
\be
M_{R}=2T_{R}S_{R}. \label{MSmarr}
\ee
Ignoring the higher orders of $\lambda$ by considering small nonextensivity ($\lambda \ll 1$ for $q\approx 1$ ), we can write down equation (\ref{MSmarr}) as
\be
M_R=\frac{r_h}{2}+\frac{\lambda \pi r_h^3}{4} \label{M_R1} 
\ee
or, explicitly in terms of the Schwarzschild radius $r_h=2M$ as
\be
M_{R}=M+2\pi\lambda M^3. \label{M_R} 
\ee
This means that the mass of the R\'enyi black hole $M_R$ is the sum of the mass of the Schwarzschild black hole plus some ``environment mass'' due to the extra term in the equation (\ref{M_R}). That is, in equation (\ref{M_R1}), $M_R$ is no longer internal energy due to the additional term. Let us notice that the equation (\ref{M_R1}) can be written as
\be
M_R=2T_{bh}S_{bh}+\frac{\lambda \pi r_h^3}{4}.
\ee 
It is shown in \cite{Promsiri:2020jga, Promsiri:2021hhv} that the nonextensive parameter $\lambda$ can be taken as a thermodynamic pressure
\be
P=\frac{3\lambda}{32}, \label{PR}
\ee
and by defining its conjugate variable as the thermodynamic volume $V=\frac{4\pi}{3}r^3_h$, we can write a consistent Smarr formula like equation \footnote{Note that, using equation (\ref{MSmarr}), the equation (\ref{MSmarr1}) can be written as $M= 2T_RS_R-2PV$, where we have used $M=2T_{bh}S_{bh}$ for the Schwarzschild black hole .} \cite{Smarr:1972kt}
\be
M_R=2T_{bh}S_{bh}+2PV.\label{MSmarr1}
\ee
Thus the R\'enyi black hole mass, $M_R$, should be interpreted as an enthalpy instead of the internal energy of
the black hole, like in the case of an AdS black hole. In this way, we can write the extended first law of thermodynamics as 
\be
dM_R=T_{bh}dS_{bh}-PdV. \label{dMR}
\ee
Within this context, interesting studies have been investigated in \cite{Promsiri:2021hhv,Promsiri:2020jga,Nakarachinda:2021jxd} for different black holes. For example, solid, and liquid phase transition, and latent heat via R\'enyi extended phase space have been studied, and black hole heat engines have also been investigated in this scenario. At least mathematically, we can say that there is an equivalent relation between the AdS black holes and R\'enyi black holes. 

In \cite{Nojiri:2021czz}, the authors analyzed thermodynamic inconsistencies while utilizing the R\'enyi black hole entropy with Hawking temperature $T_{bh}$. For instance, by applying the first law of thermodynamics, $dE_R=T_{bh}dS_R$, they found 
\be
E_R=M-\frac{4\pi\lambda M^2}{3},
\ee
where higher order terms in $\lambda$ are ignored \footnote{Here, parameter $\lambda$ corresponds to the $\alpha$ in the mentioned paper. }. The authors contend that the equation for $E_R$ differs from the black hole mass $M$, i.e., $E_R \neq M$, and that there is no physical explanation for this expression. Thus, R\'enyi entropy is not a viable option for black holes. Additionally, they claim that it conflicts with the principle of energy conservation under the scenario of spherically symmetric dust shell collapse leading to the formation of the Schwarzschild black hole. As a result, R\'enyi entropy cannot be used in conjunction with the Hawking temperature for black holes. We agree with their findings. However, there is no valid reason to employ the Hawking temperature with R\'enyi entropy. In this article, we present the thermodynamic arguments for why using Hawking temperature with R\'enyi entropy is physically unsuitable, and we analyze the corresponding R\'enyi temperature that should be used with R\'enyi entropy to prevent unphysical scenarios and inconsistencies.
We contend that the assumption that uses the Hawking temperature with the R\'enyi entropy is incorrect because, in nonextensive thermodynamics, the R\'enyi temperature $T_R=T_{eq}$ is the effective temperature associated with the equilibrium condition (\ref{beta*}). The related equilibrium entropy is the R\'enyi entropy $S_{eq}=S_{R}$. Therefore, we must utilize the R\'enyi temperature when utilizing the R\'enyi entropy and there is no correspondence between Hawking temperature and R\'enyi entropy,  
so there is no physical justification for utilizing the R\'enyi entropy while using the Hawking temperature or vice versa.
 It is worth noting that by using,  $dE_R=T_{R}dS_R$, we get $E_R=M$, which gives a consistent thermodynamic relationship between the black hole energy and mass.
Another key reason for not using the Hawking temperature $T_{bh}$ with the R\'enyi entropy $S_R$ is its inconsistency with the Legendre structure (\ref{leg1}) and (\ref{leg2}), which we shall discuss in further detail in one of the following sections. For example, to describe all thermodynamic quantities as physical variables, we must utilize physical temperature $T_{eq}=T_R$ with $S_R$ when defining free energy $F$. See, for instance, equations (\ref{leg1}) and (\ref{leg2}).  If we use $\beta=1/T_{bh}$ instead of $\beta^*=1/T_{eq}$ in $F$, then $F$ cannot be represented in physical variables, i.e., $\beta$ does not give the equilibrium condition in this case.

With the preceding arguments, we can conclude that the R\'enyi temperature and entropy have valid physical interpretations and that these quantities for a black hole are well-defined when the black hole is in equilibrium with the surroundings. This means that $T_R$ represents the physical temperature of the entire system containing a black hole embedded in some surroundings. This is simply demonstrated by the equation (\ref{M_R}), where the first term is the mass of the Schwarzschild black hole and the second term is due to work done by the environment. It is worth noting that the higher order terms in $\lambda$ are disregarded in the equation (\ref{M_R}). Additionally, the parameter $\lambda$ is somewhat related to the cosmological constant, which can be related to the pressure, giving the same extended thermodynamics for black holes. For instance, compare equations (\ref{P}), (\ref{dM}), and (\ref{M}) with equations (\ref{M_R1}), (\ref{PR}), and (\ref{dMR}). Note that, in \cite{Komatsu:2016vof}, the author used the Padmanabhan thermodynamic approach, in which the R\'enyi entropy is specified on the Hubble horizon, to obtain a term similar to the cosmological constant in the Friedmann equation. This provides yet another rationale for linking the cosmological constant and the parameter $\lambda$.

In equation (\ref{ERenyi}), we assumed that $S_q=S_{bh}$\footnote{In most of the literature for R\'enyi black hole entropy, $S_q=S_{bh}$ is substituted in $S_R$, because $S_{bh}$ is nonadditive and nonextensive. However, the problem with this assumption is that, if we rely on Bekenstein's definition, $S_{bh}$ does not follow the same composition rule for $S_q$. } and also we assume that it follows the nonadditive composition rule (\ref{S_12}). However, in \cite{Abe_2001c}, the author used a unique composition rule by using the definition of Bekenstein entropy and finding the equilibrium entropy and associated equilibrium temperature. By following \cite{Czinner:2015ena,Abe_2001}, the composition rule for black hole entropy can be written as
\be
\frac{S_{bh12}}{k_B}=\frac{S_{bh1}}{k_B}+\frac{S_{bh2}}{k_B}+2\sqrt{\frac{S_{bh1}}{k_B}}\sqrt{\frac{S_{bh2}}{k_B}}, \label{S_12bh}
\ee
 where we consider the case of two black holes, having entropies $S_{bh1}$ and $S_{bh2}$ and $S_{bh12}$ is the total entropy of composite black hole system with the inclusion of the long-range force or interactions. By maximizing the entropy (\ref{S_12bh}) with variation in total mass $\delta M_{bh12} =\delta (M_{bh1}+M_{bh2})=0$, we have the equilibrium condition
\begin{equation}
    \frac{\frac{\partial S_{bh1}}{\partial M_{bh1}}}{k_B\sqrt{\frac{S_{bh1}}{k_B}}}=\frac{\frac{\partial S_{bh2}}{\partial M_{bh2}}}{k_B\sqrt{\frac{S_{bh2}}{k_B}}}=  \beta^*,
\end{equation}
where the parameter $\beta^*$ is defined as
\be
k_B\beta^*=\frac{\beta}{\sqrt{\frac{S_{bh}}{k_B}}}=\frac{1}{T_{eq}}.
\ee 
Here, we introduced $\beta=\partial S_{bh}/\partial M_{bh}=1/k_BT_{bh}$, which is the usual inverse Hawking temperature. Now we can write $T_{eq}$
\be
T_{eq}=\frac{\sqrt{\frac{S_{bh}}{k_B}}}{k_B\beta},
\ee
and the associated equilibrium entropy can be written as 
\be
S_{eq}=2k_B\sqrt{\frac{S_{bh}}{k_B}}.
\ee 
Interestingly, the above equilibrium entropy is additive; like in the previous section, equilibrium entropy $S_R$ follows the additive rule for the general nonextensive case. For the case of the Schwarzschild black hole, the physical temperature becomes $T_{eq}=1/(4k_B \sqrt{\pi})$ and associated equilibrium entropy reads as $S_{eq}=4\pi k_B M$. This means that the equilibrium entropy is a linear function of the mass of the black hole.

In \cite{Czinner:2015ena}, the same results have been obtained by using the so-called ``formal logarithm" approach \cite{Bir__2011}, and it is shown that, within this approach, pure isolated black holes are thermodynamically stable against spherically symmetric perturbations.

\section{Composition Rule and Equilibrium Temperature for Tsallis-Cirto Black Hole Entropy}
\label{composition3}
The Legendre transform is significant in classical mechanics, statistical mechanics, and thermodynamics because it describes how information is coded in a functional form. It demonstrates how to write a function with the same information as $F(x)$, but as a function of $dF/dx$. For example, the inverse temperature $\beta = 1/ T$ is the conjugate of a system's total energy $E$. Despite this, we use the temperature $T$ in the majority of the relationships. The familiar equation
\be
F=E-TS \label{F}
\ee
which relates the Helmholtz free energy $F$ to the entropy $S$, and it hides the symmetry between $\beta$ and $E$. However, one can write the duality between them by writing the dimensionless form of (\ref{F}). In this way, Gibbs's free energy is another example. 

By following \cite{Tsallis:2012js,Tsallis:2019giw}, for a general $d$ dimensional system, the Gibbs free energy $G$ reads as
\be
G=U-TS+pV-\mu N,
\ee
where $T$, $p$, $\mu$, are the temperature, pressure, and chemical potential, and $U$, $S$, $V$, and $N$ are the internal energy, entropy, volume, and the number of particles, respectively. Here, $S$, $V$, and $N$ are the extensive variables scaling with $V=L^d$, where $L$ is the linear dimension of $d$-dimensional system, and the intensive variables $T$, $p$, and $\mu$ scaling with $L^\theta$, and finally those variables representing the energies, $G$ and $U$ scaling with $L^\epsilon$. From the above equation, it follows that
\be
\epsilon=\theta+d.
\ee 
Schwarzschild ($3+1$)-dimensional black holes have $E=M$ and $M$ scales with the length scale $L$. Since $\epsilon=1$ for this case, we obtain $\theta=1-d$ from the equation above. Let's take into account the Bekenstein entropy, which scales with $L^2$. This indicates that the temperature for a Schwarzschild ($3+1$)-dimensional black hole scales with $L^{-1}$, which is precisely true for Hawking temperature $T_{bh}$. This indicates that the quantities $S_{bh}$ and $T_{bh}$ satisfy the Legendre structure if we think of a black hole as a two-dimensional object.
Additionally, if we consider black holes as three-dimensional objects based on the aforementioned Legendre structure, the definitions of entropy and temperature alter for black holes. Tsallis and Cirto \cite{Tsallis:2012js,Tsallis:2019giw} proposed a new type of black hole entropy, and it is defined as follows:
\be
S_T=k_B\left(\frac{S_{bh}}{k_B}\right)^\delta, 
\label{Sdelta}
\ee 
where $\delta>0$ and its composition rule is given by
\begin{equation}
    S_{T12}=k_B\left[\left (\frac{S_{T1}}{k_B}\right)^{1/\delta}+\left (\frac{S_{T2}}{k_B}\right)^{1/\delta}\right]^\delta. \label{Tsallisbh}
\end{equation}
In this context, the $S_{bh}$ is additive, and $S_T$ is nonadditive. For $\delta=3/2$, $S_T$ is proportional to the volume for the case of the Schwarzschild black hole, and so it is extensive. If we consider black hole as $d=3$ dimensional system, then $S=S_{\delta=3/2}$ and $\theta=-2$, which means that $T$ must scale with $1/L^2$. The corresponding Tsallis-Cirto temperature can be written  by using $S_T$ as 
\be
T_\delta=\frac{T_{bh}}{\delta}\left(\frac{S_{bh}}{k_B}\right)^{1-\delta}, \label{Tdelta}
\ee 
which scales with $1/L^2$ for $\delta=3/2$, i.e., $T_\delta  \propto 1/M^2$, for the case of Schwarzschild black hole. 
Now using the equilibrium condition, we maximize the Tsallis-Cirto black hole entropy $S_T$, i.e., $\delta S_{T12}=0$ with the assumption that the total energy is fixed, then the equilibrium condition gives
\be
\frac{1}{k_B}\left(\frac{S_{T1}}{k_B}\right)^{\frac{1-\delta}{\delta}}\frac{\partial S_{T1}}{\partial U_1}=\frac{1}{k_B}\left(\frac{S_{T2}}{k_B}\right)^{\frac{1-\delta}{\delta}}\frac{\partial S_{T2}}{\partial U_2}=\beta^*,
\ee 
which means that the equilibrium temperature for this case can be written as
\be
T_{eq}=\frac{1}{k_B\beta^*}=T_\delta \left (\frac{S_T}{k_B}\right)^{\frac{\delta -1}{\delta}}, \label{Tphys1}
\ee 
and corresponding equilibrium entropy can be written as
\be
S_{eq}=k_B\delta \left(\frac{S_T}{k_B}\right)^{1/\delta}.\label{Seq1}
\ee
Interestingly, using the values of $S_T$ (\ref{Sdelta}) and $T_{\delta}$ (\ref{Tdelta}) in above equations (\ref{Tphys1}) and (\ref{Seq1}), we get $T_{eq}=T_{bh}/\delta$ and corresponding equilibrium entropy would be $
S_{eq}=\delta S_{bh}$. Again, in the context of the above composition rule ({\ref{Tsallisbh}}), the equilibrium entropy for this case is also additive. Note that, now the definitions of equilibrium entropy $S_{eq}=\delta S_{bh}$ and equilibrium temperature $T_{eq}=T_{bh}/\delta$ are defined in terms of $S_T$ and $T_{\delta}$. In this regard, we can say that the  Bekenstein-Hawking entropy and Hawking temperatures are the equilibrium entropy and equilibrium temperature for the nonextensive Tsallis-Cirto setup. 

Let us revisit the Legendre structure in this situation. Similarly to the situation of $S_R$, in \cite{Nojiri:2021czz}, the authors utilized the argument against the Hawking temperature $T_{bh}$ (\ref{S_bh}) associated usually with Tsallis-Cirto black hole entropy (\ref{Sdelta}) in numerous applications. In this context, they analyze that $E_T\neq M$ by applying the relation $dE_T=T_{bh}dS_T$, and therefore $T_{ bh}$ is not an appropriate choice to employ with $S_{T}$. The authors provided no reason for using $T_{bh}$ with $S_T$. This is merely an assumption; we will always obtain nonphysical results if we make inappropriate assumptions because the relationship $dE=TdS$ between temperature and entropy is required for a well-defined Legendre structure. For example, $T_{bh}$ and $S_T$ cannot be used in the thermodynamic potential $G$ because both are incompatible. To avoid unphysical outcomes, one must utilize equivalent compatible physical quantities, such as $T_{\delta}$ with $S_T$ with $dE_T=T_{\delta}dS_T$.

Formally, Tsallis-Cirto entropy as given by the formula (\ref{Sdelta}) together with (\ref{S_bh}), is the same as the Barrow entropy \cite{Barrow:2020tzx} which is defined as 
\be
S_{B} = k_B \left( \frac{A}{A_{Pl}} \right)^{1 + \frac{\Delta}{2}} ,
\ee
where $A_{Pl}$ is the Planck area, and $0 \leq \Delta \leq 1$. While comparing both definitions, we can see that \cite{ABREU2020135805}
\be
1 + \frac{\Delta}{2} = \delta ,
\ee
for both formulas to be (up to a factor) the same. The extensive limit of the nonextensive Barrow entropy is given for $\Delta = 1$, which corresponds to $\delta = 3/2$ (extensive) limit of the Tsallis-Cirto entropy. 

However, Barrow's entropy comes from purely geometrical or rather fully non-thermodynamical motivation. Shortly, the idea is to replace a black hole smooth spherical horizon with the fractal structure of spheres attached infinitely to the spherical horizon forming the so-called sphere flake, characterized by a fractal dimension $D_f$ falling into the interval $2 \leq D_f \leq 3$. This leads to an effective horizon sphere radius to be  
\be
r_{eff} =  r^{1+\Delta/2},
\ee
where $r$ is the radius of the non-fractal horizon. The horizon area is then modified accordingly 
\be
A_{eff} = 4\pi r_{eff}^2, 
\ee
and so is the (area) entropy. 

Despite that, it seems that Tsallis-Cirto thermodynamics can fully be applied to the Barrow entropy within the range of Tsallis-Cirto nonextensivity parameter $1 < \delta < 3/2$. This also means that the equilibrium temperature for Barrow entropy can also be defined, and it falls into the same formula (up to some factors) as for the Tsallis-Cirto entropy as given by (\ref{Tdelta}). Barrow entropy has recently been used in many cosmological horizon applications claiming to serve as holographic dark energy \cite{Saridakis:2020zol, MPDSalz2020,Wang:2022hun, DiGennaro:2022grw, Luciano:2022pzg, DiGennaro:2022ykp}. 

\section{Conclusions}
\label{summary}
We have explored several aspects of the nonextensive thermodynamics of black holes. In particular, by maximizing various nonextensive entropies defined on the event horizon, we have studied the equilibrium temperature for a Schwarzschild black hole and obtained the equilibrium conditions in the nonextensive setting. We have come to the conclusion that there is always an equilibrium temperature in the nonextensive setup which is different from the absolute temperature and corresponds to an additive equilibrium entropy that is different from the nonextensive one.

The primary purpose of our  study has been to determine whether the Hawking temperature was appropriate for black holes and other cosmological applications in the nonextensive scenario. In this respect, we have shown that Hawking temperature is not a consistent thermodynamic quantity to take into account while studying the nonextensive entropy of  black holes and cosmological horizons. For instance, we have shown that the Legendre structure is not valid when one associates the Hawking temperature with the R\'enyi black hole entropy and Tsallis-Cirto black hole entropy. Furthermore, we have found in the Tsallis nonextensive setup that the R\'enyi temperature was the equilibrium temperature and R\'enyi entropy was the equilibrium entropy for black holes.

In the nonextensive setup, the assumption of Bekenstein entropy as Tsallis entropy is unclear. The nonextensive nature of Bekenstein entropy provides the basis of this supposition. Bekenstein entropy, on the other hand, follows a specific nonadditive composition rule based on the entropy-area relation rather than the generic nonextensive composition rule. In this context, we have explored the equilibrium temperature by maximizing the Bekenstein entropy, which is simply a constant independent of the mass of the black hole, and the associated equilibrium entropy in this case is proportional to the mass of the black hole.

Moreover, by maximizing the Tsallis-Cirto black hole entropy, we have investigated the equilibrium temperature and have demonstrated that the Hawking temperature is the equilibrium temperature, and the Bekenstein entropy is the corresponding equilibrium entropy for such a case. A similar result is true for the case of Barrow entropy, too.

Numerous applications of non-extensive entropies in gravitation and cosmology have been made recently, yet there are still many fascinating concerns that remain unresolved. With nonextensive entropies, in particular, the Hawking temperature has been often utilized despite being inconsistent with nonextensive entropies except the standard Bekenstein entropy. Therefore, using the consistent temperature corresponding to each nonextensive entropy, we explored the consistent nonextensive thermodynamic quantities for black holes. In the future, when we get to these problems, it would be intriguing to look into the nonextensive thermodynamics of cosmological horizons with consistent nonextensive thermodynamic quantities. Similarly, it would be interesting to investigate the second law of thermodynamics and Bekenstein bound within this framework. Moreover, modified cosmology from a thermodynamic perspective with consistent thermodynamic quantities would be more interesting.

\subsection*{Acknowledgements}
The work of I.C. and M.P.D. was supported by the Polish National Science Centre grant No. DEC-2020/39/O/ST2/02323.


\end{document}